# Title: Single-crystal X-ray diffraction of BC8 carbon near terapascal pressures

**Authors:** Liang Sun[1, 2]†, Bo Chen[3]†, Zhiyu He[4], Pin Yang[1], Zanyang Guan[1], Guo Jia[4], Fan Zhang[4], Yi Yang[5], Bo Xu[5], Pei Guo[4], Yonggang Liu[1], Wei Fan[1], Tao Xi[1], Yulong Li[1], Lingbiao Meng[1], Wei Kang[5], Zhiyong Xie[4], Yuqiong Tu[4], Cheng Ji[2], Xiuguang Huang[4], Zhongjing Chen[1], Ji Yan[1], Jiamin Yang[1], Wenge Yang[2], Toshimori Sekine[2, 6], Zongqing Zhao[1]*, Jiayu Dai[3, 7]*, Ho-Kwang Mao[2]*

**Affiliations:**

[1]National Key Laboratory of Plasma Physics, Laser Fusion Research Center, Chinese Academy of Engineering Physics; Mianyang, 621900, China.

[2]Center for High Pressure Science and Technology Advanced Research; Shanghai, 201203, China.

[3]College of Science and Hunan Key Laboratory of Extreme Matter and Applications, National University of Defense Technology; Changsha, 410073, China.

[4]Shanghai Institute of Laser Plasma; Shanghai, 201800, China.

[5]HEDPS, Center for Applied Physics and Technology, and College of Engineering, Peking University; Beijing, 100871, China.

[6]Graduate School of Engineering, Osaka University; Suita, Osaka 565-0871, Japan.

[7]Hunan Research Center of the Basic Discipline for Physical States, National University of Defense Technology; Changsha, 410073, China.

*Corresponding authors. Email addresses: jydai@nudt.edu.cn (J. Dai); 498427431@qq.com (Z. Zhao); maohk@hpstar.ac.cn (H.-K. Mao).

†These authors contributed equally to this work.

**Abstract:** Whether carbon forms crystalline structures beyond diamond at terapascal pressures remains unresolved. Predicted for nearly four decades, the denser BC8 phase has lacked direct crystallographic evidence. Here we report the first direct crystallographic identification of BC8 carbon, formed as a single crystal from molten carbon during laser-driven decaying-shock compression of [100]-oriented diamond following loading above 1.1 TPa. Using *in situ* orientation-resolved single-crystal X-ray diffraction, we observe discrete Bragg spots whose spacings and angular relationships are reproduced by a common orientation matrix within each indexed BC8 reflection group, providing constraints beyond powder peak matching. Our single-crystal formation strategy and direct diffraction evidence resolve the decades-long quest for a post-diamond carbon structure, demonstrating how dynamic loading pathways can make predicted high-pressure phases experimentally accessible.

**Main Text:** Carbon underpins core problem in condensed matter physics and planetary sciences. Its well-characterized allotropes, graphite and diamond, serve as fundamental benchmark for understanding chemical bonding and structural stability across wide pressure-temperature ranges (*1–5*). In the deep interiors of Uranus, Neptune, and particularly in carbon-rich exoplanets, pressures can reach the terapascal (TPa) region, alongside temperature of several thousand kelvin (*6–10*). Under such extreme conditions, theoretical studies spanning nearly four decades have predicted BC8 carbon, a denser tetrahedrally bonded structure that becomes thermodynamically favored over diamond at ultrahigh pressure (*11–16*). Independent crystal-structure searches have further identified BC8 as the stable high-pressure carbon allotrope (*17*).

Experimentally, however, BC8 carbon has remained elusive despite sustained efforts at major facilities (*3*, *5*, *18*, *19*) and advances in in-situ X-ray diffraction (*4*, *20–22*). Ramp-compression experiments have shown that diamond can persist to pressures as high as 2 TPa without transforming to BC8 (*3*), highlighting the kinetic barriers associated with the transformation. Atomistic simulations further emphasize that pressure alone is insufficient to drive subsolidus diamond-BC8 conversion and phase formation is highly sensitive to the pressure-temperature (P-T) loading pathway (*23–25*). More recently, shock compression experiments on microcrystalline diamond reported the persistence of diamond to approximately 1 TPa and evidence of melting, again without detecting BC8 (*26*). These findings suggest that accessing BC8 carbon requires more than reaching extreme pressures, highlighting the importance of the P-T history during dynamic compression.

Our previous decaying-shock experiments on [100]-oriented single-crystal diamond established thermodynamically constraints on carbon melting and proposed BC8 formation region (*27*). As the shock decayed, the inferred P-T trajectory descended from the liquid carbon regime and exhibited a sharp kink interpreted as a possible signature of BC8 crystallization. The subsequent trajectory was consistent with the proposed BC8 melting boundary to the diamond-BC8-liquid triple point near 830 GPa and 7200 K, before approaching the diamond melting boundary, and finally entered the solid-diamond phase field. This sequence was broadly consistent with the theoretically predicted phase diagram (Fig.1b). The inferred passage through the high-temperature liquid regime suggested a possible route to BC8 formation, as enhanced atomic mobility could facilitate the structural rearrangements required for crystallization. Yet whether BC8 had formed along this pathway remained unresolved.

This high-temperature pathway also makes the crystallographic identification of BC8 particularly challenging. Although in-situ X-ray diffraction has reached terapascal pressures (*22*, *28*), rapidly evolving P-T condition can complicate diffraction measurements, while conventional polycrystalline diffraction under shock conditions may provide insufficient crystallographic information to distinguish competing high-pressure phases. Orientation-resolved SCXRD offers a more stringent approach by combining lattice-spacing information with the angular relationships among reflections, allowing candidate structures to be tested against a common crystal orientation. Here, using *in situ* orientation-resolved SCXRD, we report the first direct crystallographic identification of BC8 carbon formed from molten carbon during laser-driven decaying-shock compression of [100]-oriented single-crystal diamond. The measured reflection spacings and angular relationships identify a single BC8 crystal lattice, extending orientation-resolved crystallographic identification to near-terapascal pressures on nanosecond timescales.

**Results.**

To probe carbon near the diamond-BC8-liquid phase boundaries, we combined *in-situ* X-ray diffraction (XRD) with line-imaging VISAR at the SG-IIIp laser facility (*21*, *27*, *29*, *30*). A 1-ns, 351-nm laser pulse launched a decaying shock into a [100]-oriented single-crystal diamond, while a synchronized 1-ns quasi-monochromatic 8.4-keV Cu He-α X-rays pulse probed the compressed sample with diffraction recorded on image plates (Fig. 1A).

We used the shock pressure history to identify the conditions sampled during the X-ray exposure. In shot 242, the VISAR-constrained pressure decreased from ~1.2 to 0.8 TPa during the X-ray diffraction time. Mapping this history onto the reference Hugoniot (*5*) places the sampled conditions across the previously reported BC8 melting region near the triple point(*5*, *15*, *31*) (Fig.1B). Shot 366 used comparable drive and probe conditions and sampled an overlapping pressure range, allowing comparison between the two experiments. Each record therefore integrates contributions from regions experiencing spatially and temporally varying conditions. Further experimental details are provided in Materials and Methods in the supplementary materials.

The raw diffraction records reveal numerous discrete, localized spots alongside tantalum reference features and ambient diamond reflections (Fig. 1C and Fig. S3). The representative pattern for shot 242 contains more than ten candidate reflections. Their localized appearance is consistent with diffraction from individual crystalline domains, while their radial and azimuthal positions could provide complementary constraints for lattice indexing and crystallographic orientation.

To analyze these reflections, we mapped the image plate records onto scattering angle, 2θ, and azimuthal angle, ϕ, using the calibrated detector geometry, with tantalum reflections providing the angular calibrations (26, 29, 30). Candidate reflections were cross-checked against the raw detector images. Using provisional BC8 Miller indices, we fitted shot-specific orthorhombic lattice metrics to twelve reflections in shot 242 and seven in shot 366. We then tested whether subsets of these reflections could be described by a common crystal orientation. For each group, a single orientation matrix was fitted using signed Miller indices, and the calculated positions were compared with observations in both 2θ and ϕ (Fig. 2, A to D). Whereas 2θ primarily constrains interplanar spacings, the combined (2θ-ϕ) analysis additionally tests the relative orientation of reciprocal-lattice vectors, providing a more stringent criterion for crystallographic indexing than radial peak matching along (see Materials and Methods in supplementary materials).

In shot 242, three highlighted reflections, including $(\bar{2}00)$, (112) and (130), were approximately compatible with a common crystal orientation, yielding an angular root-mean-square (RMS) deviation of 1.64° (Fig. 2, A and B). We retained this association as a candidate group, while treating additional reflection pairs as tentative assignments. In shot 366, the three highlighted reflections, including (002), $(\overline{12}\bar{1})$and $(\overline{22}0)$, were reproduced by a common orientation matrix with a smaller angular RMS mismatch of 0.48° (Fig. 2, C and D). With the fitted lattice metrics fixed, leave-one-out tests using each pair of reflections predict the position of the omitted third reflection within the nominal 2° criterion. One additional reflection pair remains conditional. These results provide stronger orientation-based support for a common crystalline domain in shot 366.

We next examined the lattice-spacing relationships characteristic of BC8 (Fig. 3, A and B). In the ideal cubic BC8, symmetry-equivalent reflections such as (200), (020) and (002) have identical interplanar spacings. The orthorhombic metrics used in our analysis allow these spacings to split, while retaining reflections-family relationships characteristic of the BC8 framework. Across both shots, the measured spacings show patterns compatible with the BC8 {200}, {112}, {220} and

{310} families. The centers of full observed families, together with lattice-assisted estimates for incomplete families, lie close to the corresponding cubic-BC8 reference values, providing an additional consistency check on the structure assignment.

Assuming 16 carbon atoms per conventional BC8 unit cell, the fitted volumes yield nearly identical densities of 7.19 ± 0.45 g cm$^{-3}$ for shot 242 and 7.20 ± 0.46 g cm$^{-3}$ for shot 366, with uncertainties reported at ±2σ. An equal-weight combination of the two shots distribution gives a representative density of 7.19 ± 0.45 g cm$^{-3}$ (Fig. 3C). Guided by the previous melting study (25), we consider 830–1000 GPa as the inferred pressure window for BC8 crystallization during shock decay (Fig. 3D). Within this window, the measured density exceeds the liquid reference trends as expected for a denser solid coexisting with liquid. Assuming a positive entropy of fusion, this density ordering is also consistent with a positive BC8 melting-curve slope through the Clapeyron relation. As a further comparison, a density trend anchored at the previously inferred BC8 density of 6.95 g cm$^{-3}$ at 830 GPa and following the mean slope of the liquid reference trends reaches approximately 7.10 g cm$^{-3}$ at 915 GPa. This value lies within the experimental uncertainty, supporting consistency between the diffraction-derived density and the previously inferred the thermodynamic constraints.

## Discussion.

Taken together, our results establish the first direct crystallographic identification of BC8 carbon. The crystallographic assignment is supported by reflections forbidden for ideal diamond, while the agreement between lattice-derived densities from two independent experiments provides an additional consistency check. Departures from ideal cubic spacings are consistent with lattice distortion at near-terapascal pressures and temperatures of several thousand kelvin. Spatial and temporal variations in pressure, temperature and stress may contribute to these deviations, although their microscopic origin remains unresolved. Further diffraction measurements with improved temporal and spatial resolution, potentially enabled by coupling high-energy laser drivers with ultrafast X-ray free-electron laser probes (*4*, *20*), could help clarify the origin of these distortions.

The formation of BC8 carbon from molten carbon establishes a high-temperature route to this long-predicted post-diamond phase, highlighting the importance of thermodynamic pathways in accessing new crystalline structures under extreme compression. This discovery provides an experimental crystallographic benchmark for models of dense carbon relevant to planetary interiors and high-density carbon ablators in inertial confinement fusion. More broadly, this advance in extreme-condition crystallography opens new opportunities for identifying high-pressure crystalline phases in other materials. Future recovery experiments could determine whether BC8 carbon can be retained upon decompression (*32*), paving the way for investigating the predicted superhard and semiconducting properties of this long-sought carbon allotrope.

**Acknowledgments:** We acknowledge the crucial contributions from the SG-III prototype laser management, staff, and support team for the shots, especially Zhijun Li, Sai Jin, Ke Li, Ming Chen, and Dahai Yu. We also extend our gratitude to the target fabrication team, especially Weichao Tong, Hong Wei, and Yong Hu.

**Funding:**

Examples:

National Natural Science Foundation of China grant No. 12534013

National Natural Science Foundation of China grant No. 12104507

National Natural Science Foundation of China grant No. 12205276

National Natural Science Foundation of China grant No. 11975053

National Natural Science Foundation of China grant No. 41974099

Science and Technology Innovation Program of Hunan Province under grant No. 2025ZYJ001 (to J. D.).

The funders had no role in study design, data collection and analysis, decision to publish, or preparation of the manuscript.

**Author contributions:**

Conceptualization: LS, TS, HKM

Methodology: LS, BC, ZYH, WK, CJ, XGH, ZJC, JY, WGY, TS, JYD, HKM

Investigation: LS, ZYH, PY, ZYG, GJ, FZ, YY, BX, YGL, WF, TX, YLL, PG, LBM, ZYX, YQT, JMY

Formal analysis: LS, BC, ZYH, TS, WGY, JYD, HKM

Resources: ZQZ, JYD, HKM

Supervision: ZQZ, JYD, TS, WGY, HKM

Writing – original draft: LS, BC, HKM

Writing – review & editing: LS, BC, JYD, TS, WGY, HKM

All authors reviewed and approved the final manuscript.

**Competing interests:** The authors declare no competing interests.

**Data, code, and materials availability:** All data, code, and materials used in the analysis are available to any researcher for purposes of reproducing or extending the analysis.

**Supplementary Materials**

Materials and Methods

Supplementary Text

Figs. S1 to S6

Tables S1 to S5

References (*33–42*)

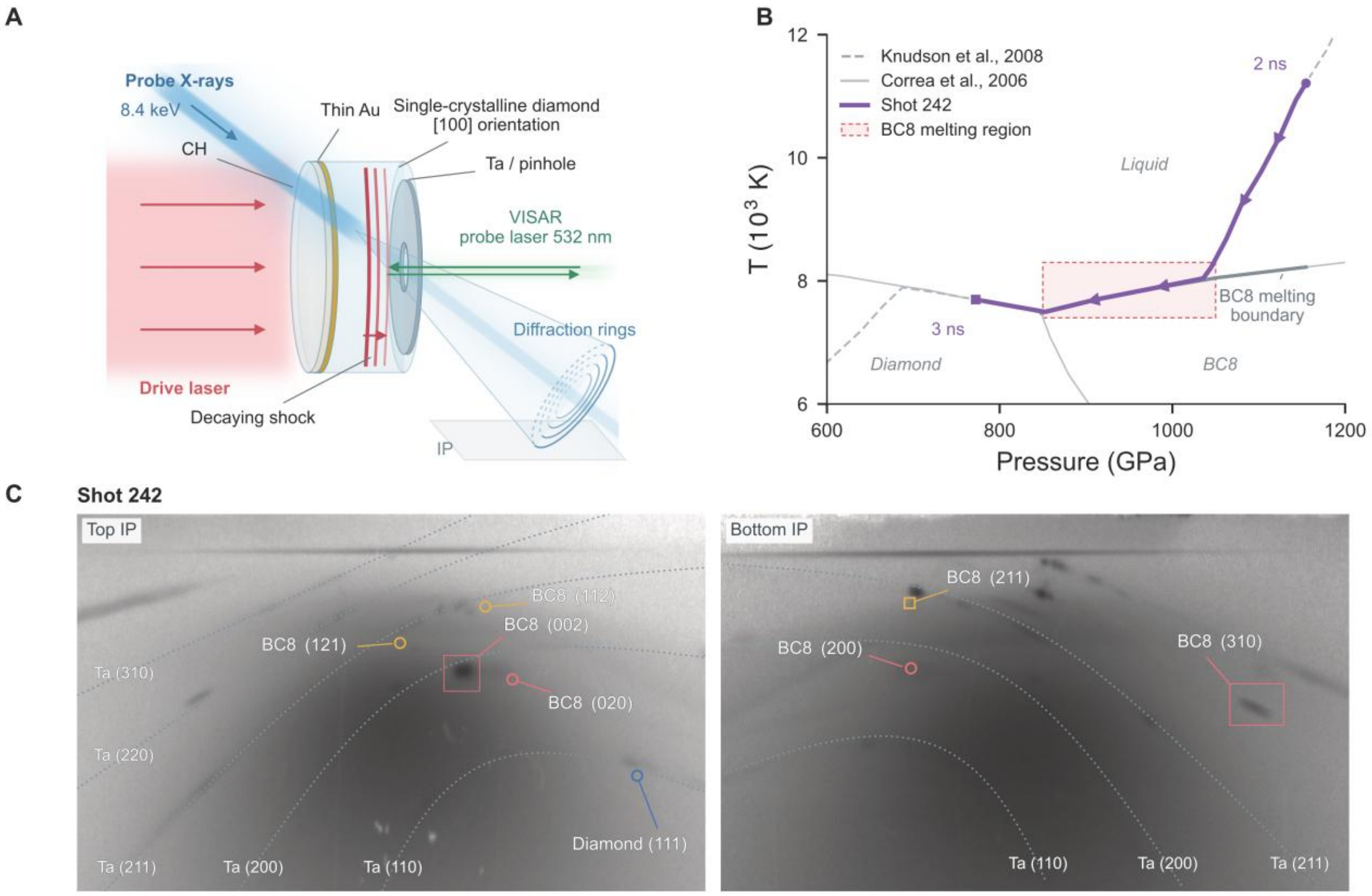

**Fig. 1 | Experimental configuration and X-ray diffraction of shocked diamond. A,** A 1-ns, ~800-J SGIII-p laser pulse drives a decaying shock into a [100]-oriented single-crystal diamond through a CH ablator and thin Au interlayer. Synchronized 8.4-keV, 1-ns X-ray pulse probes the compressed diamond with tantalum reference. A 532-nm velocity interferometer system for any reflector (VISAR) provides time-resolved velocity diagnostics. The diffracted signal is recorded in image plates (Materials and Methods). **B**, The pressure interval of 820-1200 GPa during the 2-3 ns probe interval for shot 242 projected onto the reference Hugoniot of Knudson et al. (*5*) (purple). The shaded box schematically indicates the BC8 melting region (*15*, *27*). **C,** Cropped raw X-ray diffraction records from the top and bottom image plates for shot 242. Dotted curves mark Ta and diamond reference trajectories. Colored symbols identify reflections indexed from BC8 carbon.

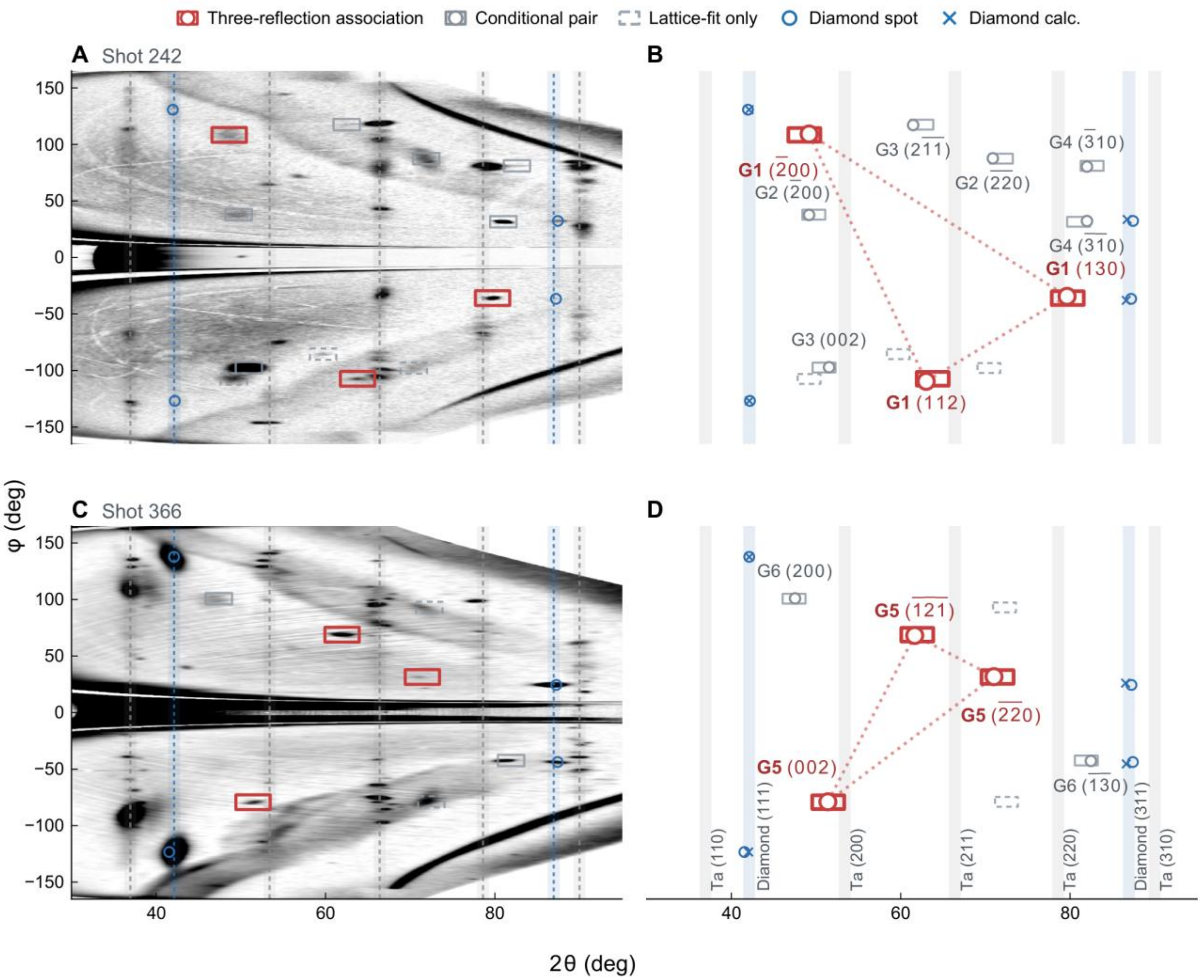

**Fig. 2. Indexed BC8 diffraction spots and orientation associations. (A and C)**, Diffraction cake images. Solid red frames mark BC8 spots belonging to three-reflection groups, while solid grey frames mark conditional two-reflection pairs and dashed grey frames mark reflections used only for lattice fitting. **(B and D)**, Measured positions (frames) and forward-calculated positions (open circles), labelled by orientation group (G1–G6) and signed Miller indices. Groups G1–G4 belong to shot 242 and G5–G6 to shot 366. Red dashed guides connect the three reflections in groups G1 and G5. Blue circles and crosses denote observed and calculated diamond reflections. Reference lines indicate Ta and diamond spacings, while shaded bands show representative ±2σ ranges from the Ta reference calibration.

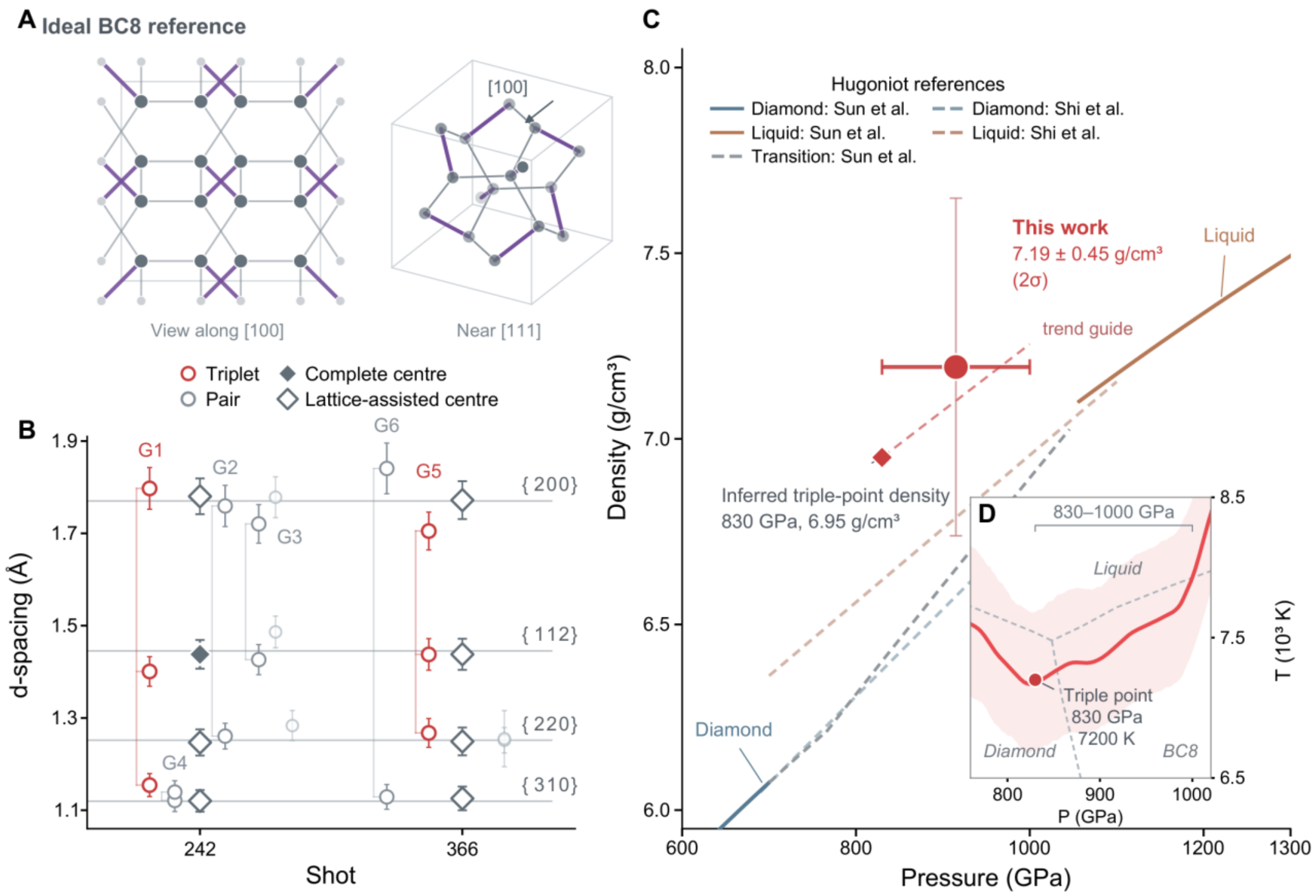

**Fig. 3 | Structure and lattice density of BC8 carbon. A,** Ideal BC8 reference structure ($Ia\bar{3}$, cI16), viewed along [100] (left) and near [111] (right). Purple and grey bonds denote one short and three long bonds per carbon atom while pale grey atoms indicate periodic neighbors and the arrow marks [100]. **B,** Indexed diffraction spacings. Red and grey open circles denote triplets and pairs connecting lines identify the G1–G6 groups in Fig. 2. Pale circles denote rest reflections. Filled and open grey diamonds mark complete and assisted family centers with errors. Horizontal lines show cubic BC8 spacings at 7.19 g cm$^{-3}$. **C,** Representative density 7.19 ± 0.45(*27*) g cm$^{-3}$, compared with Hugoniot references from Sun et al. (*27*) and Shi et al. (*24*). The vertical error indicates ±2σ of the equal-weight mixture of the two shot distributions and the horizontal span marks the adopted 830–1000 GPa pressure comparison window. The red diamond denotes the previously inferred BC8 density at triple-point conditions (830 GPa, 6.95 g cm$^{-3}$). The red dashed guide passes through this point using the mean slope of the liquid references. The grey dashed curve denotes the transition segment of the Hugoniot reference between approximately 700 and 1050 GPa. **D,** P–T results from Sun et al. (*27*), with errors and theoretical phase boundaries from Correa et al. (*15*). The red point marks the triple point (830 GPa, 7200 K). The bracket indicates the pressure window used in **C**.